\documentclass[aps,prb,reprint,twocolumn,showpacs,floatfix,superscriptaddress]{revtex4-2}

\usepackage{amssymb,amsmath,amstext}                
\usepackage{graphicx}                                               
\usepackage{epstopdf}                                               
\usepackage{color}                                                     
\usepackage{bm}                                                        
\usepackage{appendix}                                              
\usepackage[utf8]{inputenc}
\usepackage{latexsym}
\usepackage{xcolor}
\usepackage{braket}
\usepackage{ulem}
\usepackage{siunitx}
\usepackage{umoline}
\usepackage{epsfig}
\usepackage{graphics}
\usepackage{psfrag}
\usepackage{mathtools}
\usepackage[english]{babel}
\usepackage[T1]{fontenc}
\usepackage{lmodern}
\usepackage{rotating}
\usepackage{multirow}

\definecolor{lblue} {RGB}{51,71,158}
\usepackage[colorlinks=true,citecolor=blue,linkcolor=blue,urlcolor=lblue]{hyperref}

\newcommand{\new}[1]{{\color{black}{{#1}}}}

\DeclareMathOperator{\Tr}{Tr}

\begin{document}

\title{Nonergodic dynamics for an impurity interacting with bosons in tilted lattice}

\author{Pedro R. Nic\'acio Falc\~ao}
\affiliation{Szkoła Doktorska Nauk \'Scis\l{}ych i Przyrodniczych, Uniwersytet Jagiello\'nski,  \L{}ojasiewicza 11, PL-30-348 Krak\'ow, Poland}
\affiliation{Instytut Fizyki Teoretycznej, Wydzia\l{} Fizyki, Astronomii i Informatyki Stosowanej,
Uniwersytet Jagiello\'nski,  \L{}ojasiewicza 11,Krak\'ow, Poland}

\author{Jakub Zakrzewski} 
\email{jakub.zakrzewski@uj.edu.pl}
\affiliation{Instytut Fizyki Teoretycznej, Wydzia\l{} Fizyki, Astronomii i Informatyki Stosowanej,
Uniwersytet Jagiello\'nski,  \L{}ojasiewicza 11, PL-30-348 Krak\'ow, Poland}
\affiliation{Mark Kac Complex Systems Research Center, Uniwersytet Jagiello{\'n}ski, PL-30-348  Krak{\'o}w, Poland}

\date{\today}

\begin{abstract}
The fate of \new{a} single 
particle immersed in and interacting with a bath of other  particles localized in a tilted lattice is {investigated.}
{For tilt values comparable to the tunneling rate a slow-down of the dynamics is observed without, however, a clear localization of the impurity. For a large tilt and strong interactions the motion of the impurity resembles that in  the Kronig-Penney potential. The dynamics of the impurity depends on the initial distribution of majority particles. It shows delocalized dynamics for a regular, density wave like distribution and  localization if majority particles are randomly distributed.}

\end{abstract}

\maketitle

\section{Introduction}

Studies of a small impurity \new{interacting} with the medium it is immersed in have a long history.  \new{The modern formulation dates back to Landau \cite{Landau33}, who considered the motion of an electron in a dielectric crystal.} 
Due to interactions, the electron dresses in the phonon cloud forming a complex object - a polaron -  \new{named and studied} by  Solomon Pekar \cite{Pekar46, Pekar47,Pekar48,Landau48}. \new{More modern studies have considered} an impurity interacting with \new{seas} of ultra-cold bosons \cite{Cote02,Massignan05,Cucchietti06,Palzer09,Catani12,Spethmann12,Bonart12,Rath13, Fukuhara13q, Shashi14,Meinert17}
or fermions \cite{Schirotzek09, Nascimbene09, Sommer11, Kohstall12,Frohlich11, Koschorreck12,Massignan14,Yan19} . The motion of an impurity in the Bose-Einstein condensate may be considered as an example of \new{a} quantum walk \cite{Hu16a, Jorgensen16, Zoe20,Mathy12, Schecter12,Massel13, Charalambous19,Charalambous20, Lampo17, Mehboudi19,Khan21,Khan22}. 

{Recently, the study of an impurity interacting with \new{a} disordered medium received \new{significant} attention.}
Delocalization of a particle, in the presence of disorder, due to an additional coupling to \new{a} bath 
\cite{Dalessio16,Bonca17,Gopalakrishnan17,Lemut17,Bonca18,Prelovsek18} was studied in a variety of situations. A single impurity interacting with the Anderson localized medium was considered in detail in \cite{Krause21,Brighi21a, Brighi21b,Sierant23i}. {Small system studies indicated} impurity {induced} {delocalization for small disorder} \cite{Krause21}\new{,} while studies of larger systems at large disorder indicated localization at short times \cite{Brighi21a,Brighi21b}.
However, it was shown that this conclusion must be treated with caution when longer interaction times were considered \cite{Sierant23i}. The latter study extended the analysis to the quasiperiodic disorder.

The aim of the present work is to consider the dynamics of \new{an} impurity interacting with majority particles in a one-dimensional tilted lattice. This problem is disorder-free, yet recent studies have shown that known Stark localization for noninteracting particles extends also to the interacting case \cite{vanNieuwenburg19,Schulz19,Taylor20,Scherg21,Guo20,Morong21,Yao20b,Yao21,Yao21a,Doggen20s}.  Both \new{the} spectral statistics (such as gap ratio \cite{Oganesyan07,Atas13}) as well \new{the} as time dynamics reveal significant similarities between disorder induced or tilt induced localization. It is interesting, therefore, to see whether the motion of the impurity in disordered and tilted lattices also reveals common features. Some differences may be expected as it is well known
that localization may be vulnerable to external perturbations e.g. a coupling to additional phononic  \cite{Dalessio16} or bosonic baths \cite{Bonca18}. \new{Here the impurity may be considered \cite{Brighi21a,Brighi21b} to be a very small bath perturbing the Anderson or Stark localized system.} Often, in such situations one may observe subdiffusive transport
 \cite{Gopalakrishnan17,Prelovsek18}.
 
The paper is organized as follows. In Section~\ref{MM} we describe the model studied as well as define observables that we use to characterize the system. Section~\ref{SM} describes the results obtained for a relatively small system with the \new{Hamiltonian exponentiation approach} \cite{Weinberg17,Weinberg19} that allows us to analyze long-time behavior while
in Section~\ref{LA} we consider larger systems using a variational approach based on tensor networks. \new{Both approaches are compared briefly in the Appendix.} Section~\ref{RA} considers
the role of random positional disorder in the initial state for 
the dynamics - here we consider larger systems only. We conclude in Section~\ref{CO}.

\section{The model and methods} 
\label{MM}

We consider two species of hard-core bosons residing in a common one-dimensional lattice with open boundary conditions described by the Hamiltonian: 
\begin{eqnarray}
 H&=& J\sum_{i=1}^{L-1} (\hat{d}_i^\dagger \hat{d}_{i+1} + H.c.) + \sum_{i=1}^L h_i \hat{n}_{d,i}  \cr
 &+& J\sum_{i=1}^{L-1} (\hat{c}_i^\dagger \hat{c}_{i+1} + H.c.) +U   \sum_{i=1}^L \hat{n}_{c,i}\hat{n}_{d,i} ,
 \label{eq:hamb}
\end{eqnarray}
where $J$ denotes the tunneling amplitude assumed, for simplicity, to be the same for both $c$ and $d$ species (later on we fix the units with $J=\hbar=1$). $\hat{d}_i, \hat{d}_i^\dagger$ ($\hat{c}_i, \hat{c}_i^\dagger$) are the annihilation and creation operators for $d$ ($c$) bosons at site $i$, {while $\hat{n}_{d,i}$ ($\hat{n}_{c,i}$) {are their occupation number operator at site i.}  The interaction strength between the two species is characterized by $U$, while the on-site chemical potential acting on the $d$ particles is characterized by $h_i$}. A single-$c$ particle does not feel any potential and it would move freely for $U=0$. We shall study how the interactions with $d$
particles affect the $c$-impurity dynamics and how the presence of the impurity {affects} the $d$ particles.

A similar model \new{was} introduced in \cite{Brighi21a,Brighi21b}\new{,} where it was assumed that  $h_i$ is randomly and uniformly distributed, $h_i\in[-W,W]$\new{, resulting} in Anderson localization for $d$-particles alone.
While short-time study \cite{Brighi21a,Brighi21b,Brighi23} hinted at the appearance of many-body localization (MBL) in the system for sufficiently large interactions $U$  between $d$ particles and $c$ and appropriately chosen disorder amplitude $W$,
the study of longer times showed that MBL-like behavior is rather a transient effect and the impurity spreads sub-diffusively in the system\new{,} leading also to the slow delocalization of $d$-species \cite{Sierant23i}.

\new{
Let us underline that the hard-core bosons model maps to spin 1/2 system (with an empty site corresponding to the spin-down state while the occupied site to the spin-up state or vice versa). Those in turn, via Jordan-Wigner transformation, may be transformed into spinless fermions. Thus the model, \eqref{eq:hamb} is quite general. The hard-core assumption
arises in cold atomic physics when the on-site interaction among particles is huge. Then the double (or higher) occupancy on a given site is energetically very costly and the motion of particles may be restricted to a single occupancy subspace. Such an approach is widely utilized in studies of dynamical and ground state properties of, e.g. dipolar systems \cite{Varney12,Wu20,Li21}. Also the impurity localization problem in hard-core boson models has been studied in the prior works \cite{Brighi21a,Brighi21b,Sierant23i}.}

In our study we shall use a similar set of parameters as  \cite{Brighi21a,Brighi21b,Sierant23i} concentrating, in particular, on $U=12$ case, but we shall replace the random uniform disorder with a constant
tilt of the lattice $h_i=Fi$. Also, we use similar approximate numerical approach to study time dynamics, i.e. the time-dependent variational principle (TDVP) technique with matrix product states. Specifically,
we follow the prescription given in \cite{Paeckel19,Goto19} and use a combination of 2-site and 1-site codes, for details see \cite{Sierant23i}.  For small system sizes, up to $L=24$, we use also the exact time propagation as supplied \new{by} the QuSpin code \cite{Weinberg17,Weinberg19} comparing its results with that of TDVP propagation for similar system sizes \new{in the Appendix}.

Note that the Hamiltonian, \eqref{eq:hamb}, \new{is} particle-hole \new{symmetric} so the motion of a single $c$-impurity in the presence of $\overline n$-filling of $d$-bosons will be the same as for $1-\overline n$
filling. For this reason, we shall consider two filling cases.  Inspired by \cite{Brighi21a,Brighi21b,Sierant23i} we consider first the 1/3 filling of $d$-bosons prepared in a Fock, charge density wave-like state,  i.e.,  $|0,1,0,0,1,0,0,...\rangle$. Those particles, in the presence of 
a constant lattice tilt, undergo Wannier-Stark localization. Next, we discuss the 1/2 filling with $d$-particles being in $|1,0,1,0..\rangle$ initial state that leads to stronger interactions for the same $U$. Finally, we consider the situation when $d$-particles are  randomly distributed introducing the positional disorder. In all cases, a single $c$ boson is placed in the middle of the chain at the site $i_0$ which is initially empty. Its spreading in the lattice is affected by the interaction with $d$-particles. 

\new{Note that it is also possible to place the impurity in the site already occupied by the $d$-particle. This, however, shifts the energy of the system by a large value of $U$, so the system is strongly perturbed by the impurity. The tunneling from such a state is strongly nonresonant slowing down considerably the initial time evolution.}

 \begin{figure}
    \centering
    \includegraphics[width = 0.49 \textwidth]{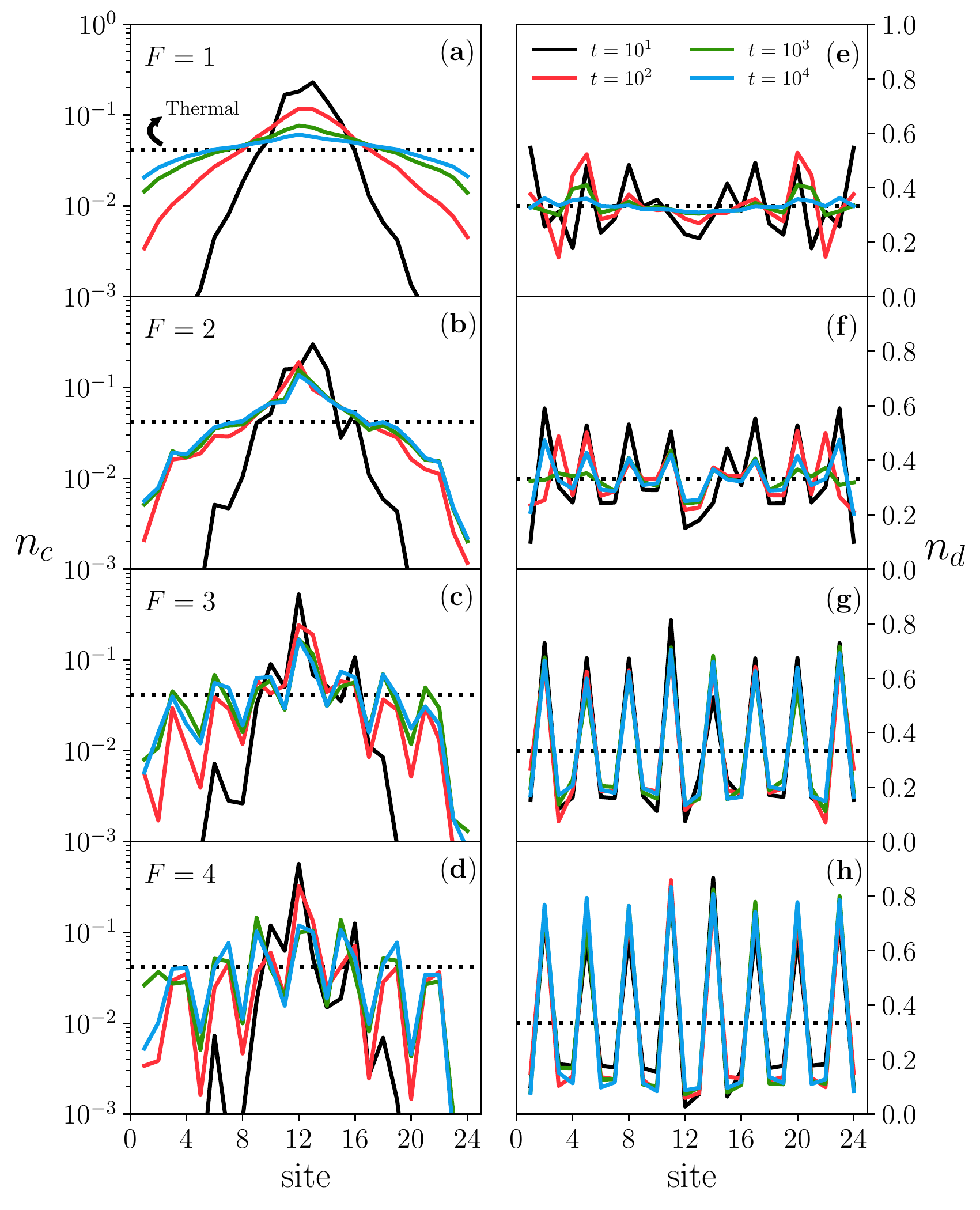}
  \caption{Impurity density, $n_c$ (left) and tilted hard-core bosons density, $n_d$ (right) for different values of the tilt $F$, as indicated in the panels. The initial state of tilted bosons is $|010010..010\rangle$,  while the impurity is placed in an initially empty central site of $L=24$ chain (total Hilbert space dimension being  $1.76\cdot10^7$). The interaction strength is $U=12$. Dotted lines indicate equilibrium occupations.
  }\label{fig:comp}
 \vspace{-0.4cm}
\end{figure} 

The full {dimension of the Hilbert space for our problem is }
\begin{equation}
    \mathrm{dim} \mathcal{H} = L \times {L\choose N_d},
    \label{Hilbert_dim}
\end{equation}
with the first term corresponding to the {Hilbert space dimension of the impurity $c$} and the second term associated with the {number of } $d$-type hard-core bosons (or fermions), {$N_d$}. The filling is, therefore, $\overline n=N_d/L$.
In the following, we shall present the ``exact'' numerical data for $L$ around 20, which corresponds to the Hilbert space dimension of the order of a few millions.  For larger systems, of the order of $L=60$, we use an approximate TDVP algorithm based on matrix product states (MPS) (for details see the Appendix and \cite{Sierant23i}).

We consider the following observables: The impurity density, {$n_{c,i}(t)=\langle \hat{n}_{c,i}(t) \rangle$}, which reveals a possible spreading of the $c$-boson through the chain, and the density of the {background bosons} affected by the tilt of the lattice, $n_{d,i}(t)=\langle \hat{n}_{d,i}(t) \rangle$. We find that
a useful information is provided by the mean square deviation (MSD) for $c$ boson position defined as
\begin{equation}
{\mathrm{MSD}(t)= \sum_i^L n_{c,i}(i-i_0)^2 -\left(\sum_i^L n_{c,i}(i-i_0)\right)^2.}
\label{dis}
\end{equation}
 The MSD is similar to the quantity studied in \cite{Bera17,Weiner19} in the context of MBL.  
 The diffusive spread is indicated by a linear growth MSD$(t)\propto t$. A saturation after an initial growth is a hallmark of Anderson localization. It is believed \cite{Bera17,Weiner19} that the growth of MSD in a genuine many-body interacting system in the MBL phase is slower than any power in time and persists for a long time. 

Additionally, we consider the time correlation function of $d$ particles defined as:
\begin{equation}
    C_d(t, L_b)= \sum_{l=L_b+1}^{L-L_b} (n_{d,l}(t)-\overline n)(n_{d,l}(0)-\overline n),
    \label{corr}
\end{equation}
where $\overline n$, the mean density which for our initial state, is, typically, $\overline n=1/3$. $L_b$ denotes the number of sites on both edges of the chain removed from the correlation function. To avoid boundary effects we take $L_b=3$. If the memory of the initial state is lost, $n_{d,l}(t)$ should reach $\overline n$ at
all sides \new{leading to vanishing $C_d(t, L_b)$}.

 We analyze also the entropy of entanglement by splitting the system on the $i$-bond into two subsystems $A$ and $B$ with $A$ containing the first $i$ sites. This allows us, in principle, to test whether the logarithmic entropy growth, considered as another feature of MBL appears in the system studied. \new{ Let us define the density matrix for the subsystem $A$ as $\rho_A(t)=\Tr_B |\psi(t)\rangle\langle \psi(t)|$ where $|\psi(t)\rangle$ is the quantum state of the system at a given time $t$. Then the entanglement entropy across this bond reads 
 \begin{equation}
     \mathcal{S}=-\Tr[\rho_A \ln \rho_A] = - \sum_k \rho_k\ln \rho_k
 \end{equation}
 where $\rho_k$ are the squares of the Schmidt basis coefficients fulfilling $\sum_k \rho_k$=1 (see e.g. \cite{karol}).
  Further, we can split the entanglement entropy $\mathcal{S}$ into two parts \cite{Lukin19}: 
\begin{equation}
    \mathcal{S} = \mathcal{S}_N + \mathcal{S}_C
    \label{Ent_entropy}
\end{equation}
where  $\mathcal{S}_N$ denotes the number (classical) entropy describing particle number fluctuations between $A$ and $B$
subsystems and $\mathcal{S}_C$ denotes the configurational (quantum) entropy (describing various particles configurations in $A$ and $B$).  Denote by $p_n$
the probability of occupying the $n$-particle sector in $A$ and by $\rho^{(n)}$ the corresponding block of the density matrix $\rho_A= \sum p_n \rho^{(n)}$. Then \cite{Lukin19}}:
\begin{align}
    & \mathcal{S}_C = -\sum_n p_{n}\Tr(\rho^{(n)}\ln \rho^{(n)}),\label{Conf_entropy}\\
    & \mathcal{S}_N = -\sum_n p_{n}\ln p_n.
      \label{Num_entropy}
\end{align}.

\section{Small system sizes}
\label{SM}

We begin our analysis with systems of small sizes amenable to an exact time propagation. They allow us also to reach longer times
of evolution.
We consider 1/3 filling first. Fig.~\ref{fig:comp} presents the time dynamics of both the impurity and the background bosons in the tilted lattices for different values of the tilt, $F$. We assume the interaction of $c$ and $d$ species to be quite high, $U=12$ as in \cite{Brighi21a,Brighi21b}. {For a small tilt ($F=1$), the impurity decays from the initial site and tends to spread over the whole lattice, suggesting a slow delocalization towards its thermal behavior. This tendency to an ergodic behavior can also be seen by looking at the density profile of the $d$-type bosons, where they lose completely the memory of their initial configuration, "melting" to $n_{d,i} = \overline n$. 
An interesting scenario emerges when we increase the magnitude of the tilt to $F=2$. Despite an initial spread, the impurity density seems to freeze in a specific configuration thus suggesting some sort of localization. This localized behavior is also reflected in the background bosons, which do not \new{completely lose} the memory of their initial configuration, even when the system is evolved for a very long time (up to $10^{4}J^{-1}$).}

 \begin{figure}
    \includegraphics[width = 0.49\textwidth]{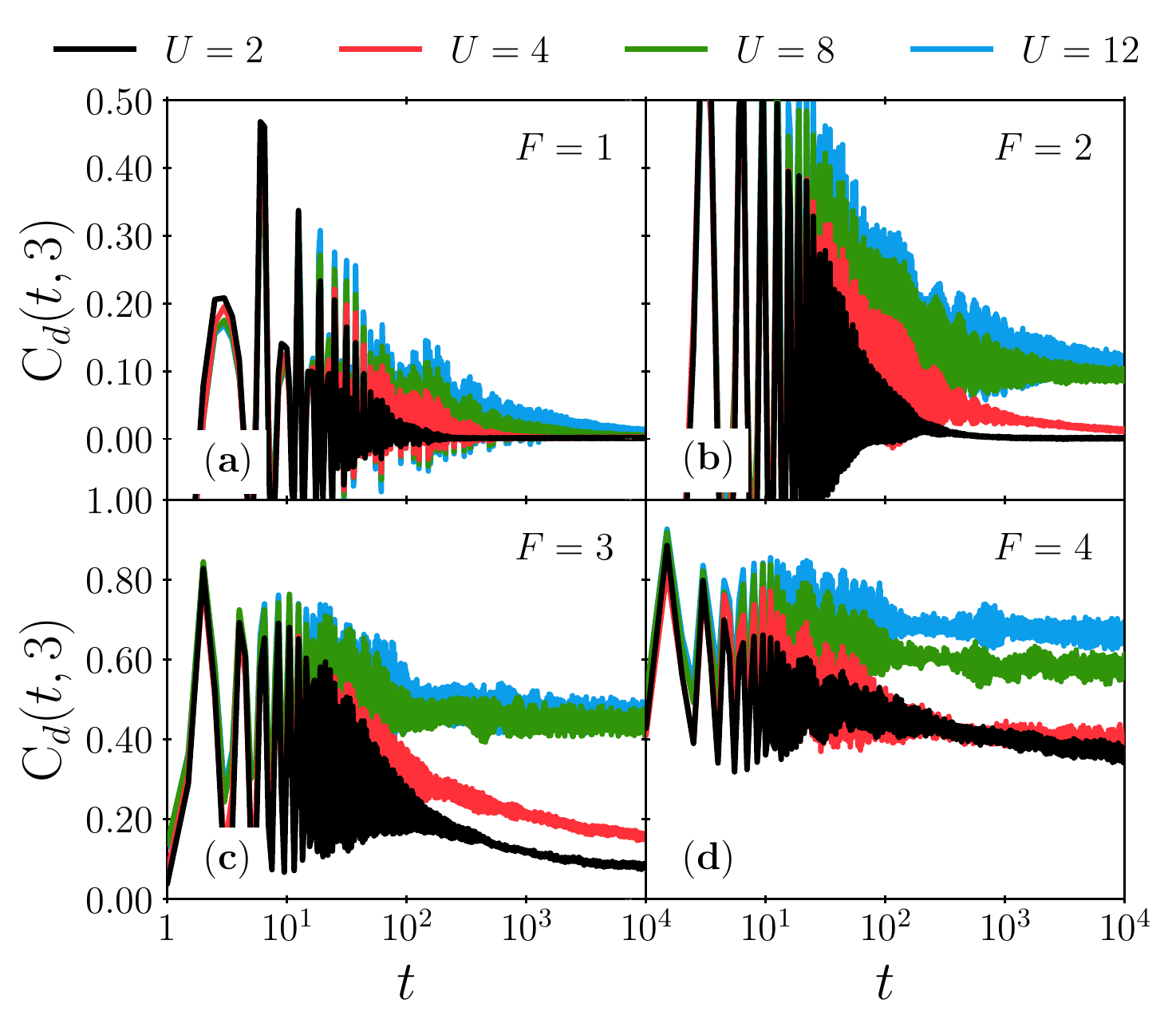}
  \caption{Time dependence of the correlation function for $d$-particles for different values of the interaction strength $U$,
  (as indicated above in the figure) and different tilt values, (indicated in the panels)
  and $L=21$ chain shows that only for strong interactions the motion remains non-ergodic on time scales considered. \new{To avoid boundary effects, we drop} $L_b=3$ sites from both edges in the calculation of $C_d(t, L_b)$.
  }\label{fig:L24corr}
\end{figure} 

This is confirmed by the time dependence of the correlation function, \eqref{corr} shown in Fig.~\ref{fig:L24corr}. Here we show the system dynamics for different interaction strengths, $U$. {For smaller $U$ values, the correlations, after initial oscillations, decay fast to zero indicating ergodic behavior.} This is to be contrasted with large $U$, in particular the $U=12$ case, where the oscillations of the correlation function decay slower and the asymptotic values (not reached even for long $10^4$ times considered) seem to be nonzero, at least for the $F>2$ cases. The oscillations are the remnants of Bloch oscillations, which would dominate the dynamics of $d$ particles for $U=0$. 

\begin{figure}
    \centering
    \includegraphics[width = 0.49\textwidth]{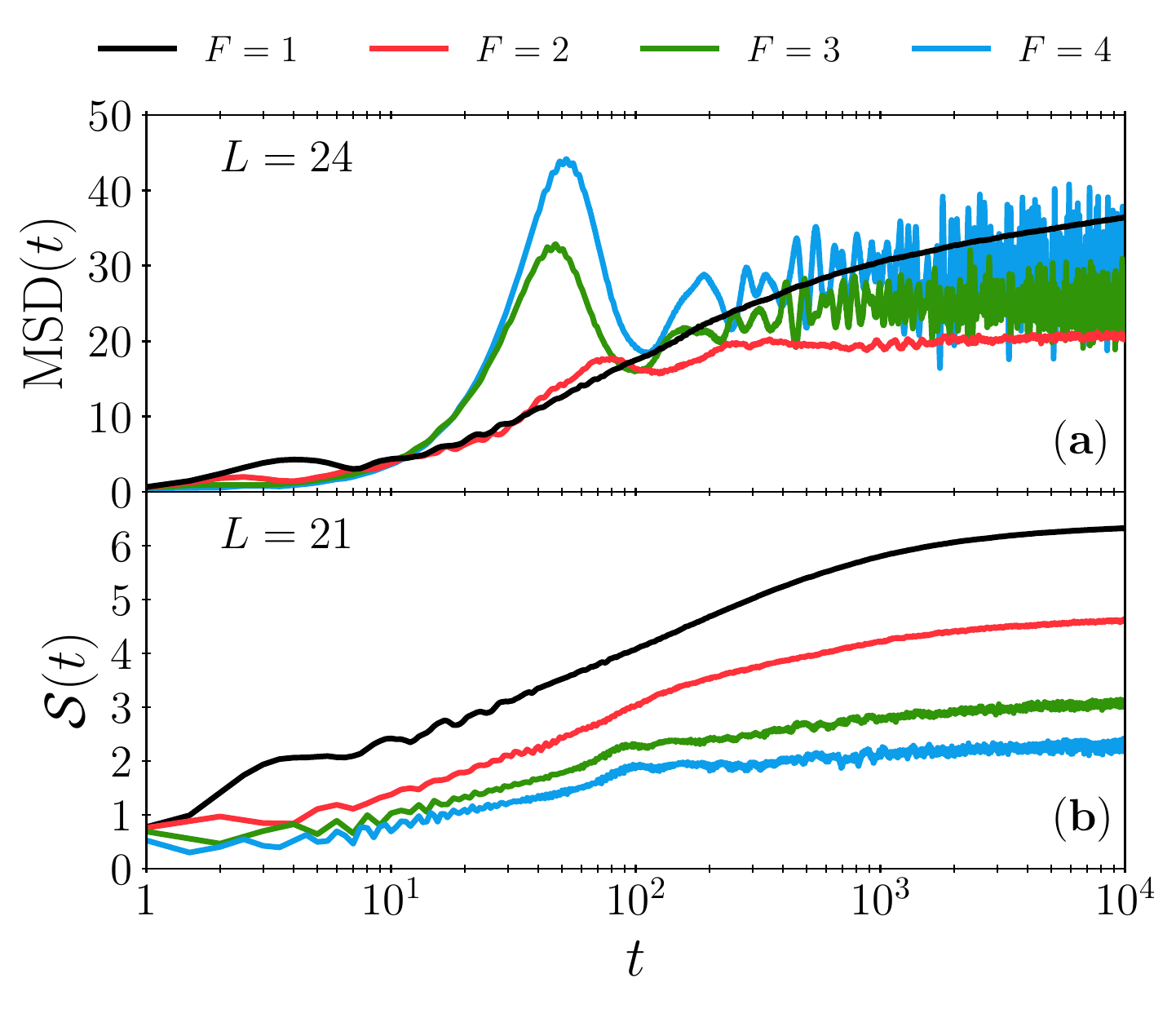}
  \caption{(a) Mean square deviation MSD($t$) time dependence for $c$ particle distribution for $U=12$ and different values of $F$ \new{for system size $L=24$} (b) the time dynamics of the entanglement entropy in the middle of the chain
  for slightly smaller system size, $L=21$.
  }\label{fig:L24MSD} 
\end{figure}

The interesting, nonergodic dynamics \new{therefore occurs} for strong interactions, from now on we shall consider mostly the $U=12$ case.

The hat-like behavior of the impurity density with the maximum at its original position, {particularly visible for $F=2$},  suggests some sort of memory of the initial \new{position of the $c$-particle}. It is accompanied by a strong modification of the background $d$-boson distribution in this central position. This behavior persists 
at quite large times (\new{up to} $10^4 J^{-1}$) where both $n_{c,i}$ and $n_{d,i}$ seem to keep their shape being extremely weakly dependent on time.

To provide a more quantitative picture {of the impurity dynamics},
Fig.~\ref{fig:L24MSD} shows the growth of MSD, Eq.~\eqref{dis}, for different $F$ values and strong interactions. This growth necessarily reflects the trends that may be invoked from Fig.~\ref{fig:comp}. For small $F=1$ the wavepacket apparently spreads quite fast, the spread slows down in time but does not stabilize on the time scale considered. Markedly different
is the growth of MSD for $F=2$
case. \new{The initial growth saturates around $t=200$ and then remains practically constant with possibly a very slow growth.} On the other hand, for still bigger tilt, $F=3,4$
one observes initially a fast superdiffusive growth ($\sim t^\alpha$ with $\alpha> 1$) of MSD which then stops growing but rather undergoes significant oscillations. By comparing the results obtained for different system sizes\new{, not shown,} one
may verify that the time  when the first maximum of MSD is formed is roughly proportional to the system size $L$ and is related to the moment when the front of the wavepacket reaches the edges of the system. At roughly the same time the entanglement entropy, Fig.~\ref{fig:L24MSD}b), 
growth changes from a steep growth to a slow logarithmic-like behavior for $F=3,4$ while for smaller $F$ values studied,
the growth is substantially faster. \new{Let us note that we report calculation of $S$ for a slightly smaller system size, $L=21$, since the corresponding calculations for $L=24$ would require excessive computer time. Still, we show MSD in Fig.~\ref{fig:L24MSD}a) for the largest $L$ we could handle. }

\new{Let us leave aside for a moment the fast initial growth of MSD in the initial $t\in[0,100]$ period.} Clearly, the dynamics for $t>100$ and larger values of $F$ studied 
 shows features expected for MBL: the distribution of $c$ as well as $d$ particles stabilizes in out-of-equilibrium configurations with close to exponential distribution for $c$ particle centered at its initial position (and revealing fast oscillations as visible from MSD fast changes at later times). Still, the mean value of MSD changes very slowly, following the MBL signature of 
Ref.\cite{Bera17,Weiner19}. The entanglement entropy shows a slow logarithmic-like growth in agreement with  another hallmark of MBL \cite{Znidaric08,Bardarson12}.

\begin{figure}
    \centering
    \includegraphics[width = 0.49\textwidth]{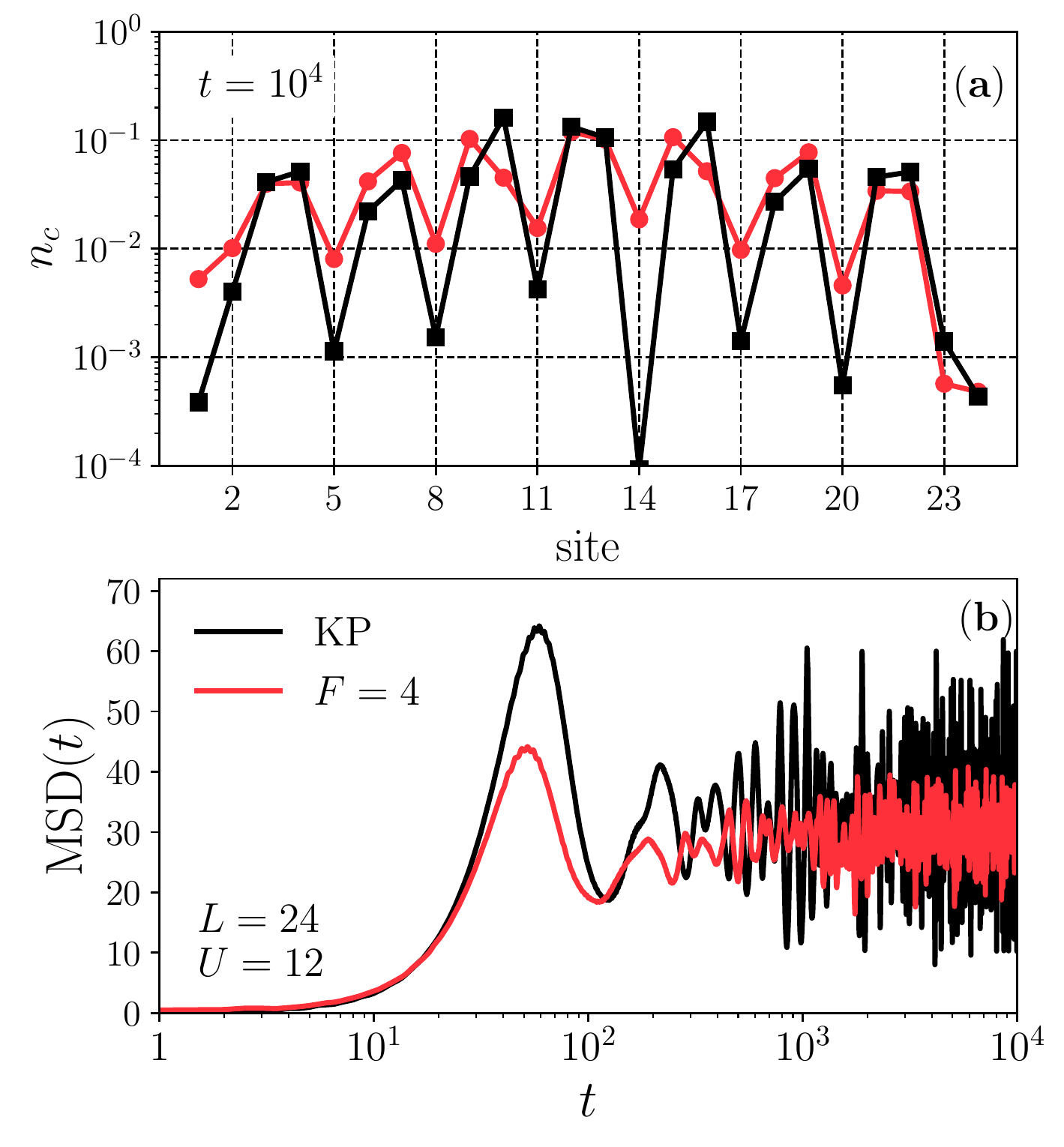}
  \caption{Comparison of the exact dynamics and that generated by a static Kronig-Penney (KP) potential of amplitude $U=12$, \new{Eq.~\ref{eq:KP}}.
  }\label{fig:L24KP} 
\end{figure}

Can we somehow provide a physical understanding of this dynamics? \new{An} interesting picture emerges \new{noting the fact} that $d$ particles for such large $F$ values are practically  localized at their initial positions. Suppose that they are strictly immobile. Then, they provide a periodic Kronig-Penney (KP) $\delta$-type potential \cite{Kittel} of amplitude $U$
for the motion of $c$-particle. \new{Then the effective Hamiltonian governing the motion of the $c$ impurity is:
\begin{equation}
    H_{\mathrm eff}= 
     J\sum_{i=1}^{L-1} (\hat{c}_i^\dagger \hat{c}_{i+1} + H.c.) +U   \sum_{i\in {\cal O}} \hat{n}_{c,i},
     \label{eq:KP}
\end{equation}
}
\new{where ${\cal O}$ is a set of sites occupied initially by $d$ particles, i.e., ${\cal O}= \{2,5,8,..\}$.}

Such an asymptotic picture may \new{only be approximate,} as it does not depend on $F$ value (\new{still under the assumption that $F$ is large}). However, already in Fig.~\ref{fig:L24MSD}, the time dependence of MSD for $F=3$ and $F=4$ looks quite similar, Fig.~\ref{fig:L24KP} compares $F=4$ case with the dynamics \new{obtained with the Hamiltonian, \eqref{eq:KP}}. The latter is a strictly periodic problem soluble by extended Bloch waves. Still, the fact that
the \new{potential acts only on every third site} has pronounced consequences for the impurity dynamics. Notably, the occupation of sites remains nonuniform, with sites where \new{the} (repulsive) potential $U$ is present being less occupied than other sites, observe the black squares in Fig.~\ref{fig:L24KP}a). This may be seen also in MSD($t$) which undergoes fast oscillations (after the initial ballistic growth) around a value ($t \sim 30$).

\new{Let us note that apparently the effective Hamiltonian explains the character of the fast growth of MSD for small times. It represents ballistic motion in the approximate KP potential.}

Interestingly, in the presence of interactions, \new{the} MSD remains very similar to that obtained for the noninteracting case while the entanglement entropy dramatically changes. For a single particle in an immobile background, the entanglement entropy grows\new{,}{reaching the value $\ln(2)$}. This contribution comes solely from the number entropy, with the value emphasising that the particle is either on the left or on the right of a given bond.  In
the interaction case the entanglement entropy  grows fast initially until times of the order of hundred when the first bump in MSD occurs. Then the growth slows down and remains logarithmic up to the longest times ($10^4$) considered as seen in Fig.~\ref{fig:L24MSD}b). Thus, despite similarity with the non-interacting KP case in particle distributions, we observe here a hallmark of MBL - the logarithmic growth of the entanglement entropy.

\begin{figure}
    \includegraphics[width = 0.45\textwidth]{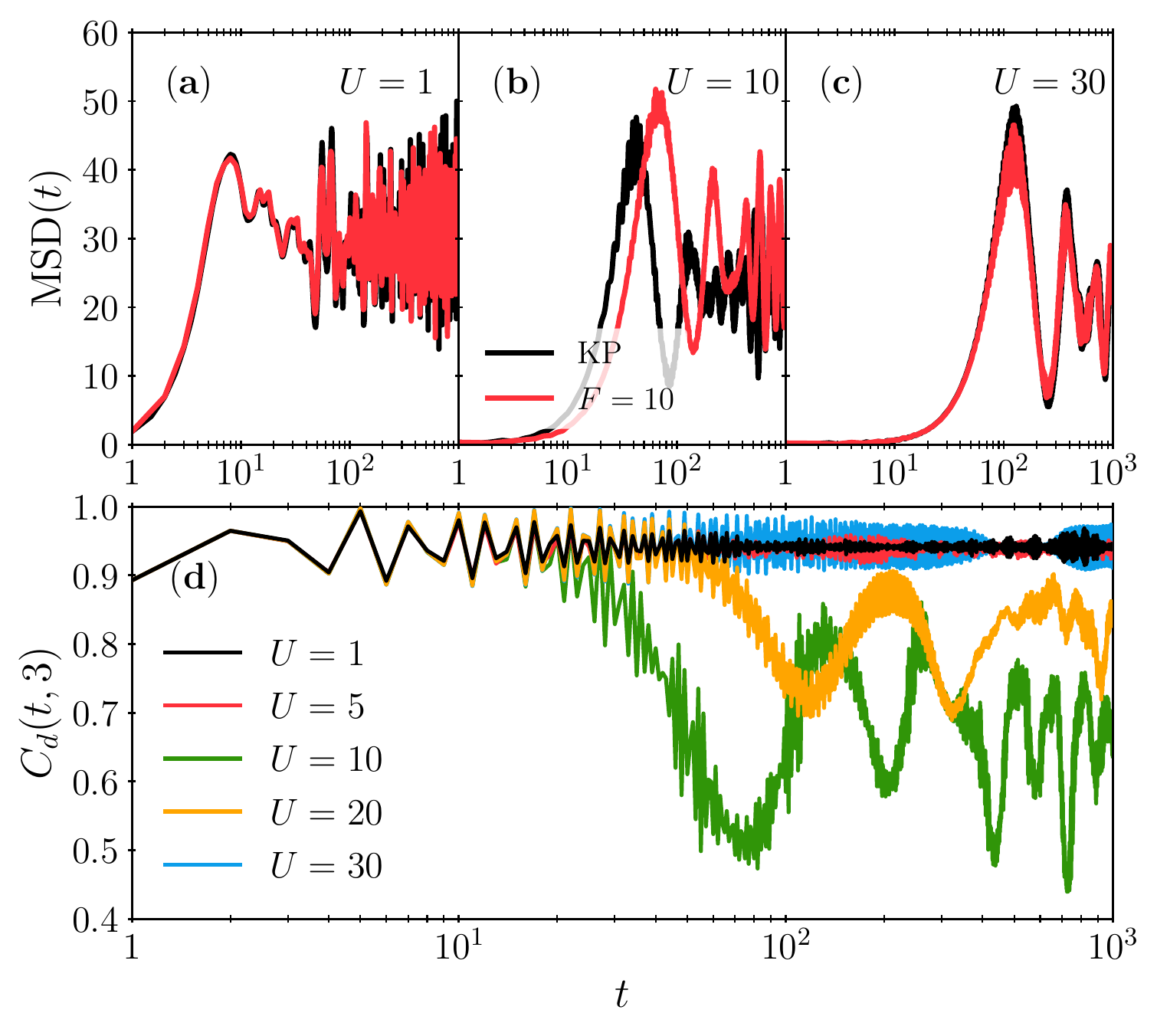}
  \caption{
  \new{Comparison of the exact dynamics with the KP approximate one for strong tilt, $F=10$ and different $U$ values. 
  Observe that the deviations from KP approximation as well as a partial decorrelation of $d$-particles is the strongest for resonance $F=U$ when the difference in energies between neighboring sites due to the tilt may be compensated by the interaction energy. Here the system size $L=21$.}
  }\label{fig:F10} 
\end{figure}

\new{The validity of the KP potential approximation depends also on $U$ value. 
Consider a really strong $F=10$ case. Fig~\ref{fig:F10} shows  MSD and the correlation function of $d$-particles for different $U$ values. For both weak and strong interactions one observes that the KP approximate description works well for the impurity, also correlations between $d$-particles are close to their original value. At resonance, $F=U$ (and in its vicinity) this picture partially breaks. When $F=U$ the difference of energies between neighboring sites due to the tilt is equal the interaction energy present if $c$ and $d$ particles meet at a given site. Thus locally the tunneling becomes resonant. This is, however, a local process, and does not extend over many sites thus leading to a partial decorrelation of $d$ particles only. Large $U\gg F$ leads to almost impenetrable barriers of the KP potential. That slows down the dynamics as seen by comparing time dynamics in top row of Fig.~\ref{fig:F10}. At the same time a full correlation between $d$-particles is preserved. Note that the time scale in Fig.~\ref{fig:F10} is up to $t=10^3$ only
as computer simulations for large $F$ and $U$ values are time costly.}

\begin{figure}
    \centering
    \includegraphics[width = 0.49\textwidth]{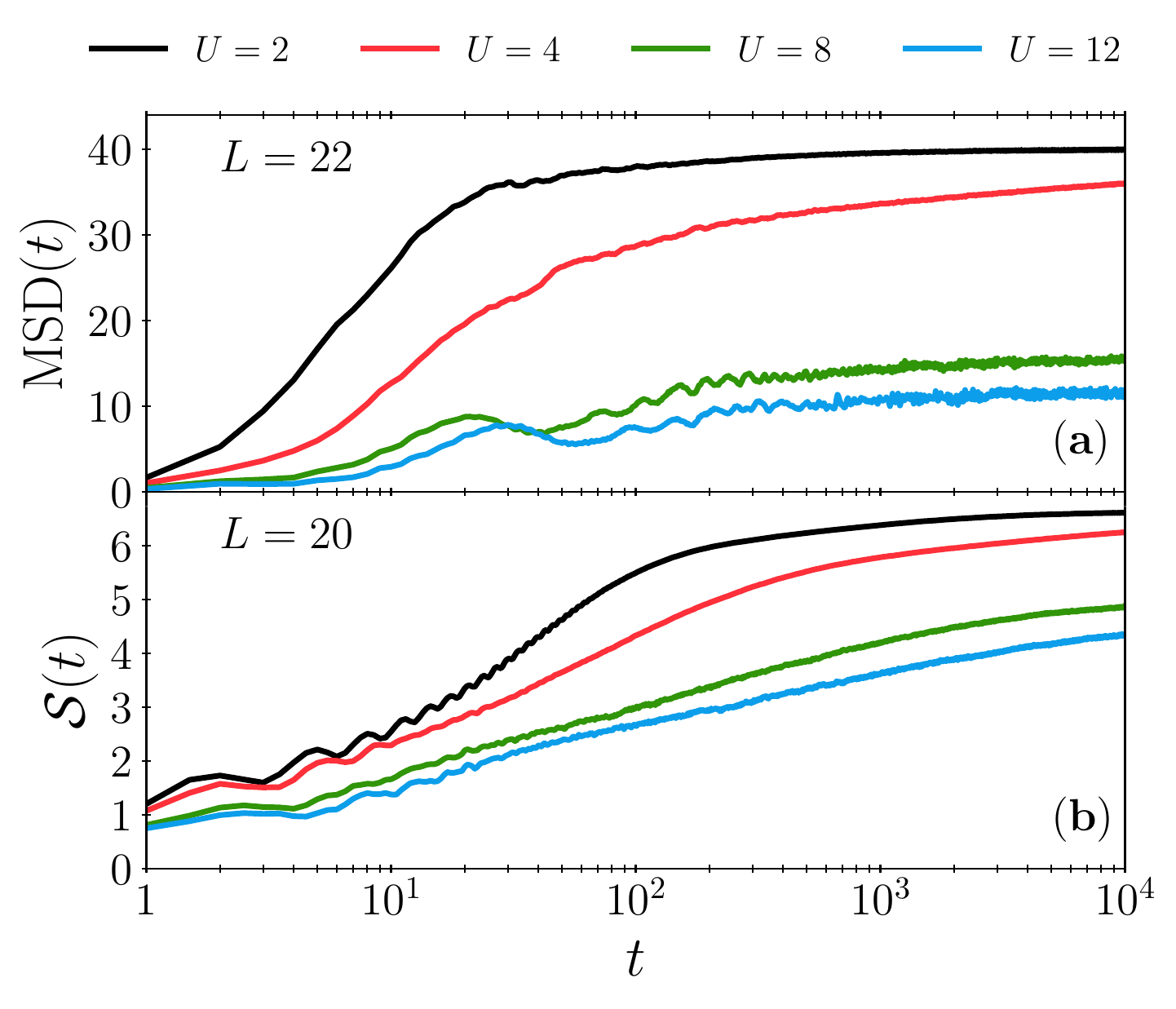}
  \caption{The time dynamics at half filling. (a) Mean square deviation MSD($t$) time dependence for $c$ particle distribution for $F=2$ and different values of $U$. (b) The corresponding time dynamics of the entanglement entropy in the middle of the chain for a slightly smaller size $L=20$.
  }\label{fig:Half1} 
\end{figure}

\begin{figure}[t!]
    \centering
    \includegraphics[width=0.49 \textwidth]{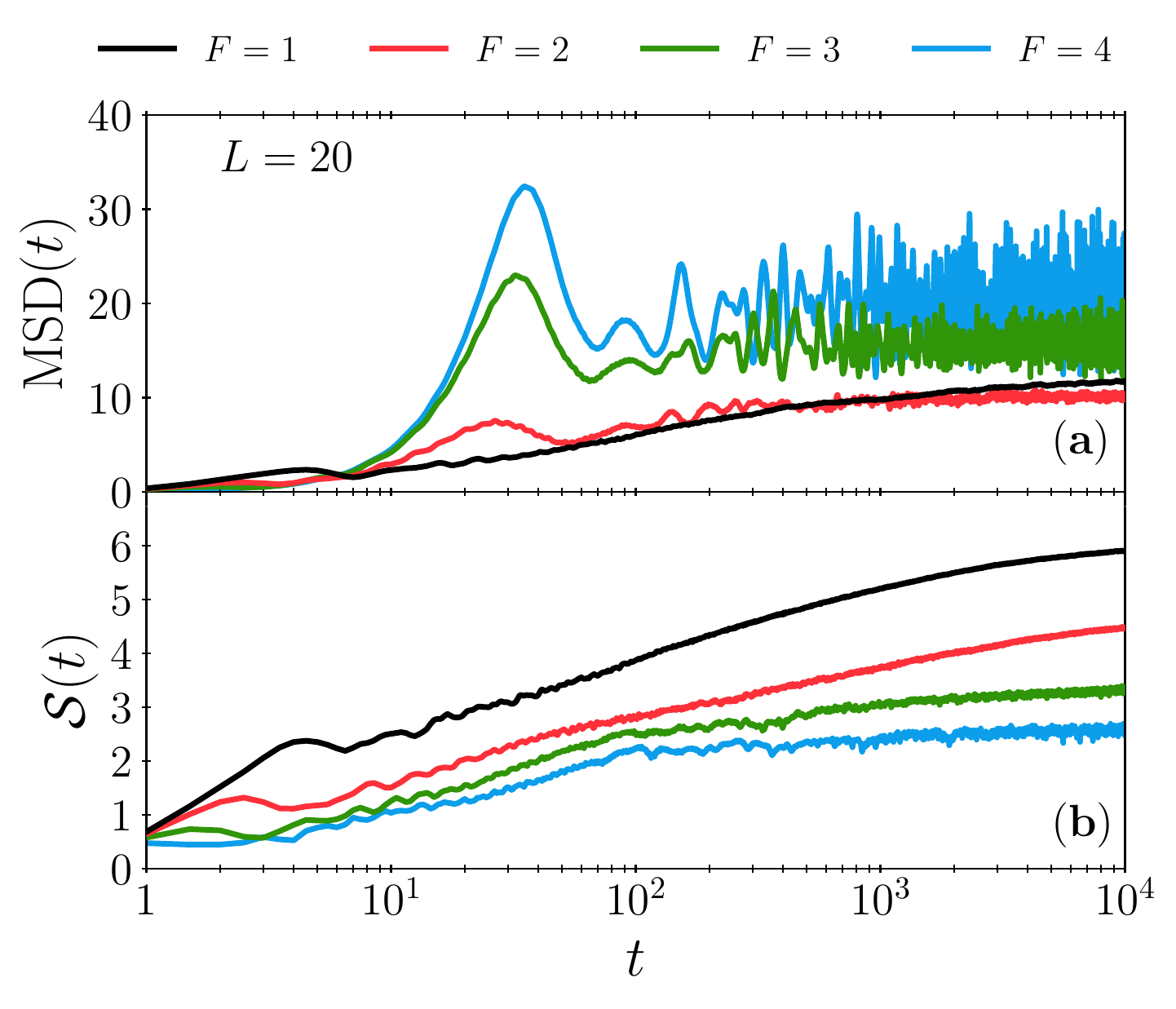}
    \caption{Long time evolution of $\mathrm{MSD}$ in the highly interactive regime ($U=12$) for different tilts. For small tilts, the mean-square deviation shows a really slow growth followed by a saturation at a small value. When $F$ is large, $\mathrm{MSD}$ experiences a fast ramp followed by oscillations around a fixed value.}
    \label{MSD_F_half}
\end{figure}

Consider now, briefly, the ``optimal'' density of $d$-particles, i.e.
$\overline n=1/2$. As an initial state we take a \new{product} state $|101010...\rangle$.
We observe a similar behavior to that for $\overline n=1/3$. Fig.~\ref{fig:Half1} presents the MSD(t) and the corresponding entanglement entropy  for different interaction strengths $U$ and $F=2$. For smaller $U$ values a fast growth of both MSD and $S$ indicates a lack of localization, while the data for $U=12$, on the other side, show a very modest value of MSD with an almost logarithmic growth
of the entanglement entropy. Note that while the data are obtained for $L=22$ with the Hilbert space dimension of $1.55\cdot 10^7$, for the entanglement entropy dynamics we report the data for $L=20$ with the corresponding dimension of $3.7\cdot10^6$.

We inspect again the $U=12$ case, comparing now MSD for different $F$ values shown in Fig.~\ref{MSD_F_half}. While for large $F$ we observe a similar behavior as for the smaller density - the MSD time dependence shows great similarity with results for KP potential with frozen $d$-particles, even for $F=2$, MSD(t) grows slowly being accompanied by a similarly slow, logarithmic growth of the entanglement entropy.
Thus increased effective interactions for half-filling favor clearly MBL also for smaller tilt values. In  particular, already for $F=2$ we observe a (sub)logarithmic growth of the entanglement entropy. Importantly for that $F$ value MSD grows also very slowly and is limited to low values indicating freezing of the corresponding time dynamics.

\section{Large sizes}
\label{LA}

 \begin{figure}
    \includegraphics[width = 0.49\textwidth]{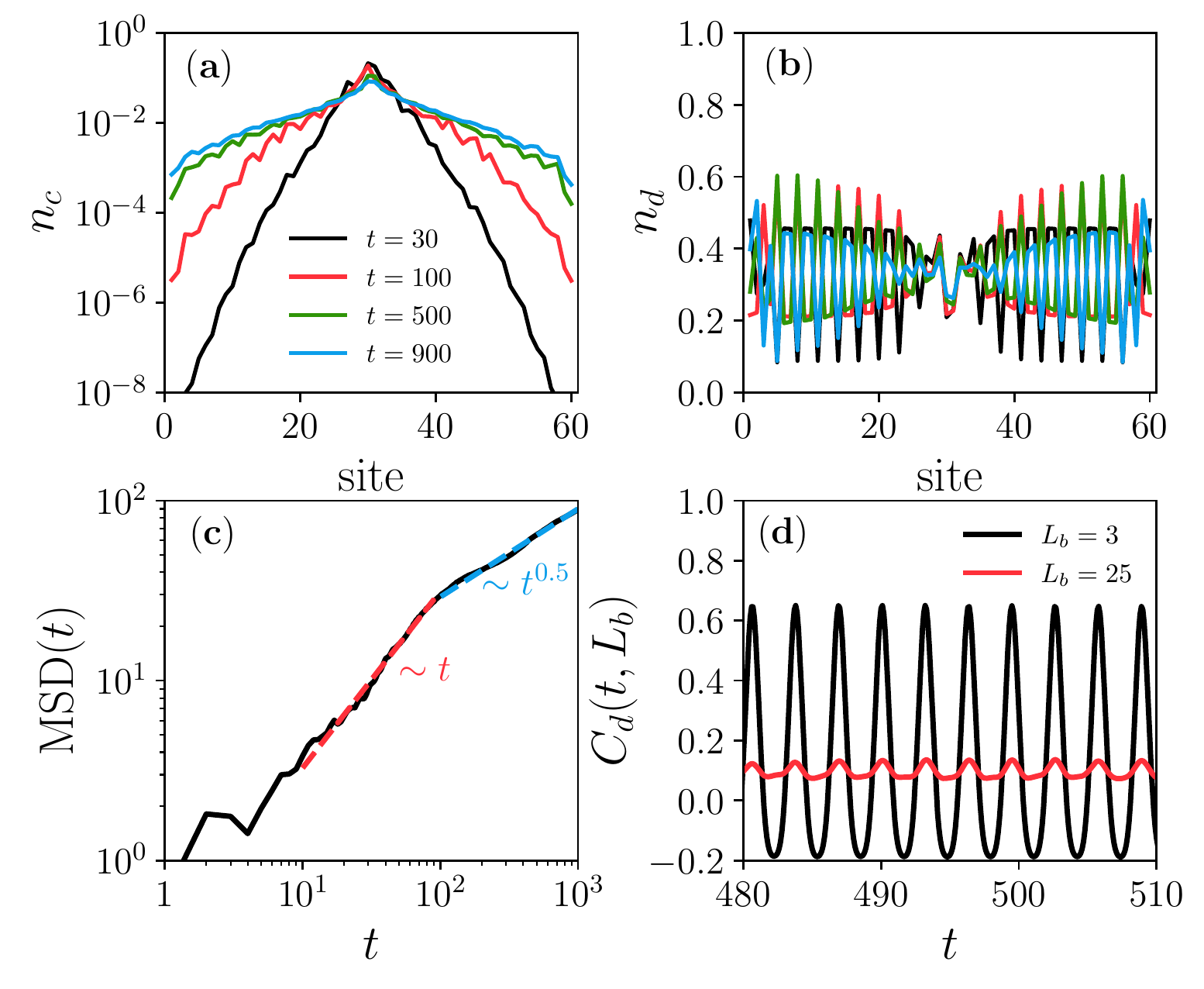}
  \caption{Time dynamics for $L=60$ chain. The interaction strength is $U=12$ and the tilt $F=2$. The $c$-impurity spreads significantly initially as seen from $n_c$ distribution (a) and MSD (c). The latter plot reveals that even at long time the spreading is significant. Panel (b) shows the distribution of $d$ particles, delocalized in the middle of the chain but keeping a partial memory of the initial state at larger distances from the center. The correlation function (d) shows significant Bloch oscillations that occurs mainly in the outer regions. The red line traces weak local correlations in the central region.
  }\label{fig:L60dens}
\end{figure} 

\begin{figure}
    \centering
    \includegraphics[width = 0.49\textwidth]{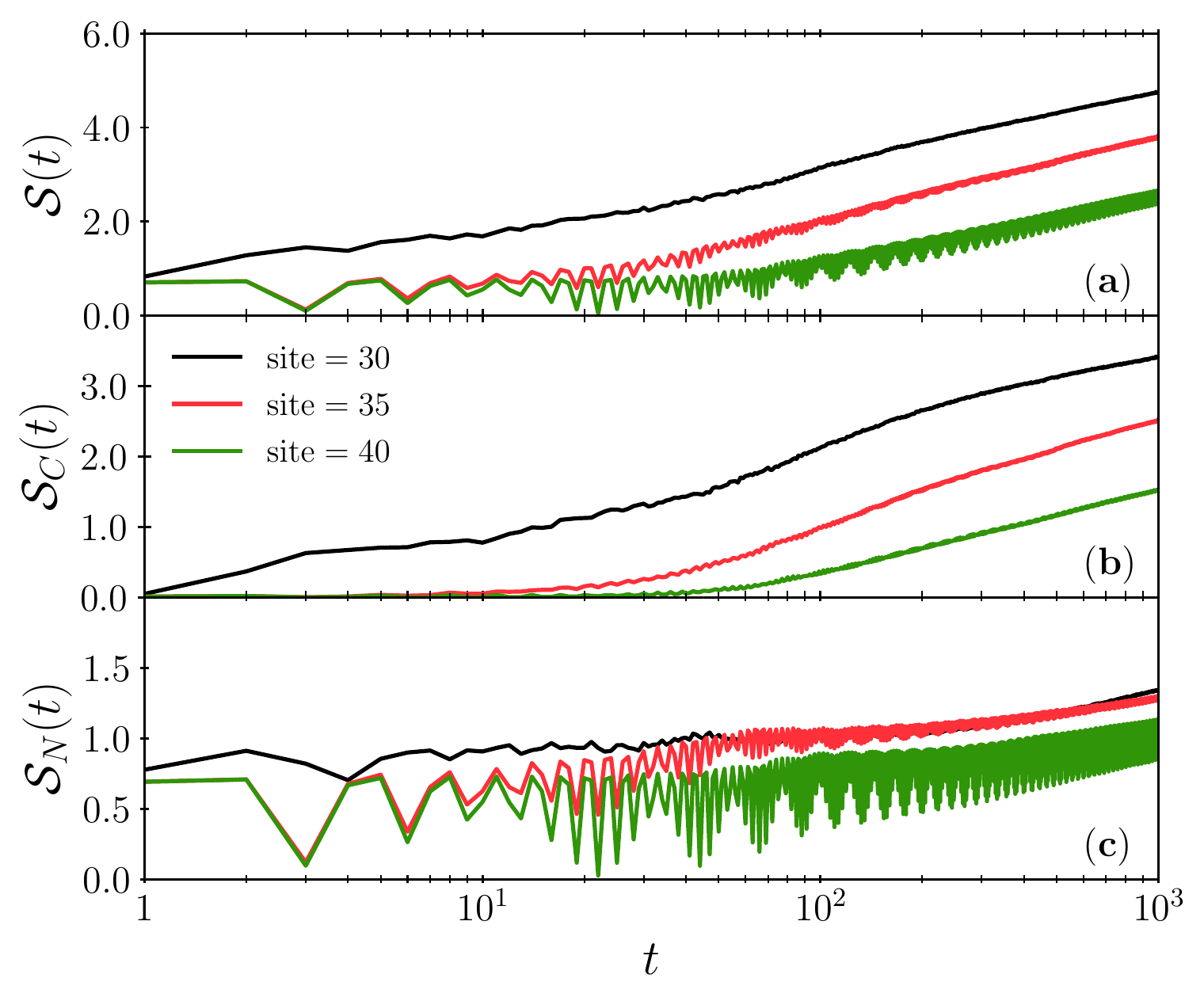}
  \caption{The time dynamics of the full entanglement entropy, top, panel (a), of its quantum (configurational) part $S_C$ (b) and
  the classical number entropy, $S_N$, bottom (c), for the same parameters as in Fig.~\ref{fig:L60dens} i.e. $U=12$, $F=2$. Black, red and green lines give entropies at the center of the chain at site $i_0=30$, at $i_0+5$ and $i_0+10$, respectively. Observe that the entropy growth is delayed for bonds further from the center.
  The central bond entropy seems to be growing slightly faster than logarithmically.
  }\label{fig:L60ent} 
\end{figure} 
Larger system sizes require different numerical methods to treat, as exact time evolution requires exponential memory resources. 
Fortunately, we can use for this purpose a thoroughly tested TDVP algorithm \cite{Paeckel19,Goto19} which we used in a number of situations involving the dynamics in nonergodic regimes \cite{Chanda20,Chanda20t,Chanda20c,Chanda20m,Sierant22,Sierant23i}. We use a combination of single-site and double-site
algorithm with dynamically grown auxiliary space dimension up to typically $\chi=384$. Tests on a number of cases \new{(see the Appendix for more details)} and in particular smaller $\chi$, show that we obtain typically below \new{$1\%$ error} for local variables (occupation of sites) with few percent for the entanglement entropy and MSD up to times $t=1000$ considered. The typical time step used was $dt=0.05/J$. 

Consider first the one-third filling by $d$ particles.
The remnants of the impurity localization observed in Fig.~\ref{fig:comp} could be at first attributed to the small system size, but TDVP evolution of much larger $L=60$ system (for a considerably shorter time) confirms that quasi-stationary long-time impurity distribution seems to carry the information about its initial position - compare Fig.~\ref{fig:L60dens}. The particle seems to spread quite fast for short times with an apparent slowdown around $t\approx 100$ as observed by the time evolution of MSD - panel Fig.~\ref{fig:L60dens}c). While initial growth seems {diffusive} for later times, $t>100$\new{,} one may fit MSD with a sub-diffusive growth, $t^\alpha$ with  $\alpha\approx 0.5$. This explains the small difference between $c$-particle density distributions at late times (as shown in panel a) of Fig.~\ref{fig:L60dens}.

 \begin{figure}
    \includegraphics[width = 0.49\textwidth]{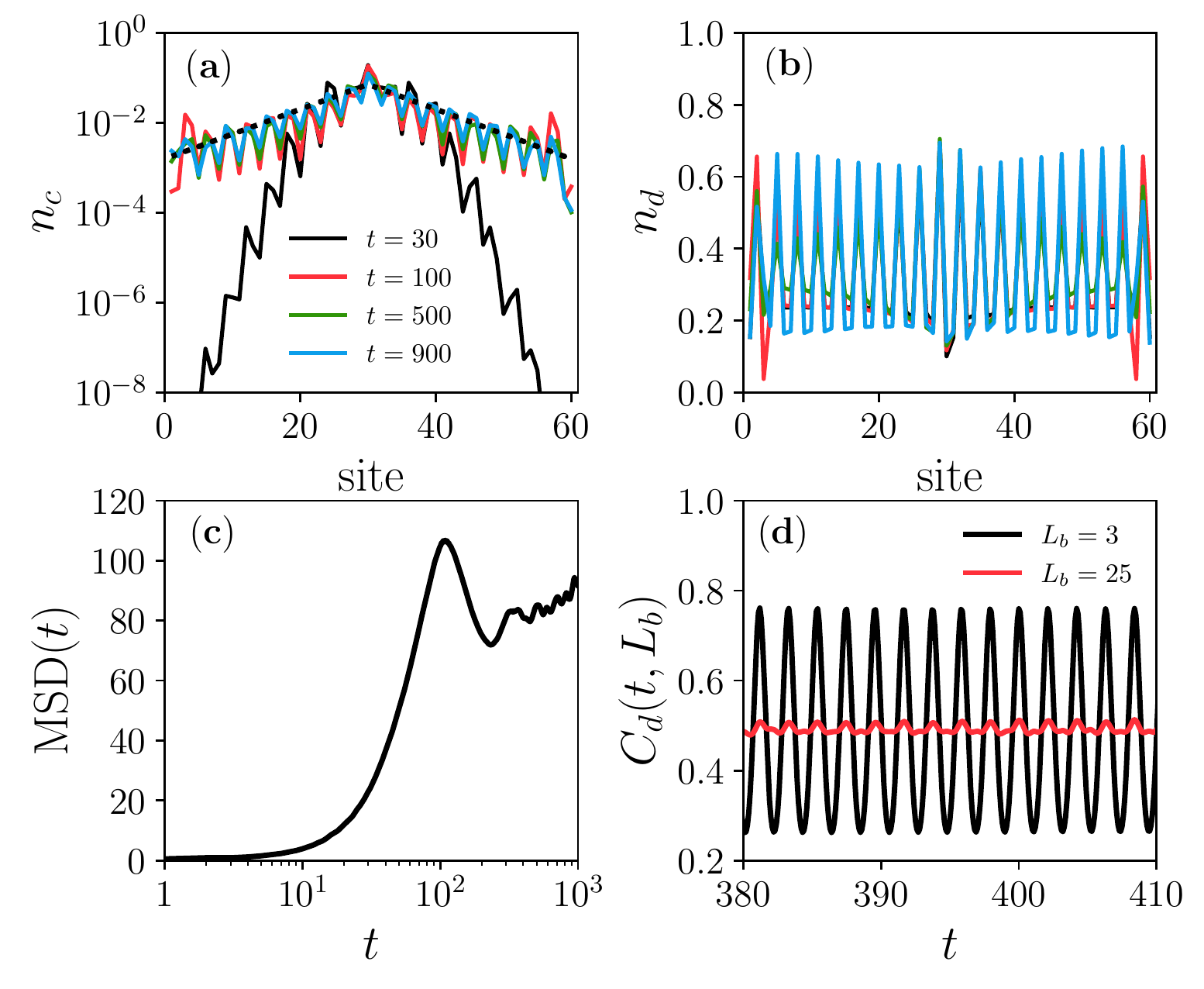}
  \caption{Same as Fig.~\ref{fig:L60dens} but for a larger tilt $F=3$. \new{Dashed lines in panel (a) represents an attempt at exponential fit of the density $\sim \exp(-B|i-30|)$ with $B=0.13$.}
  }\label{fig:L60F3}
\end{figure} 

 \begin{figure*}[t!]
    \includegraphics[width = 0.9\textwidth]{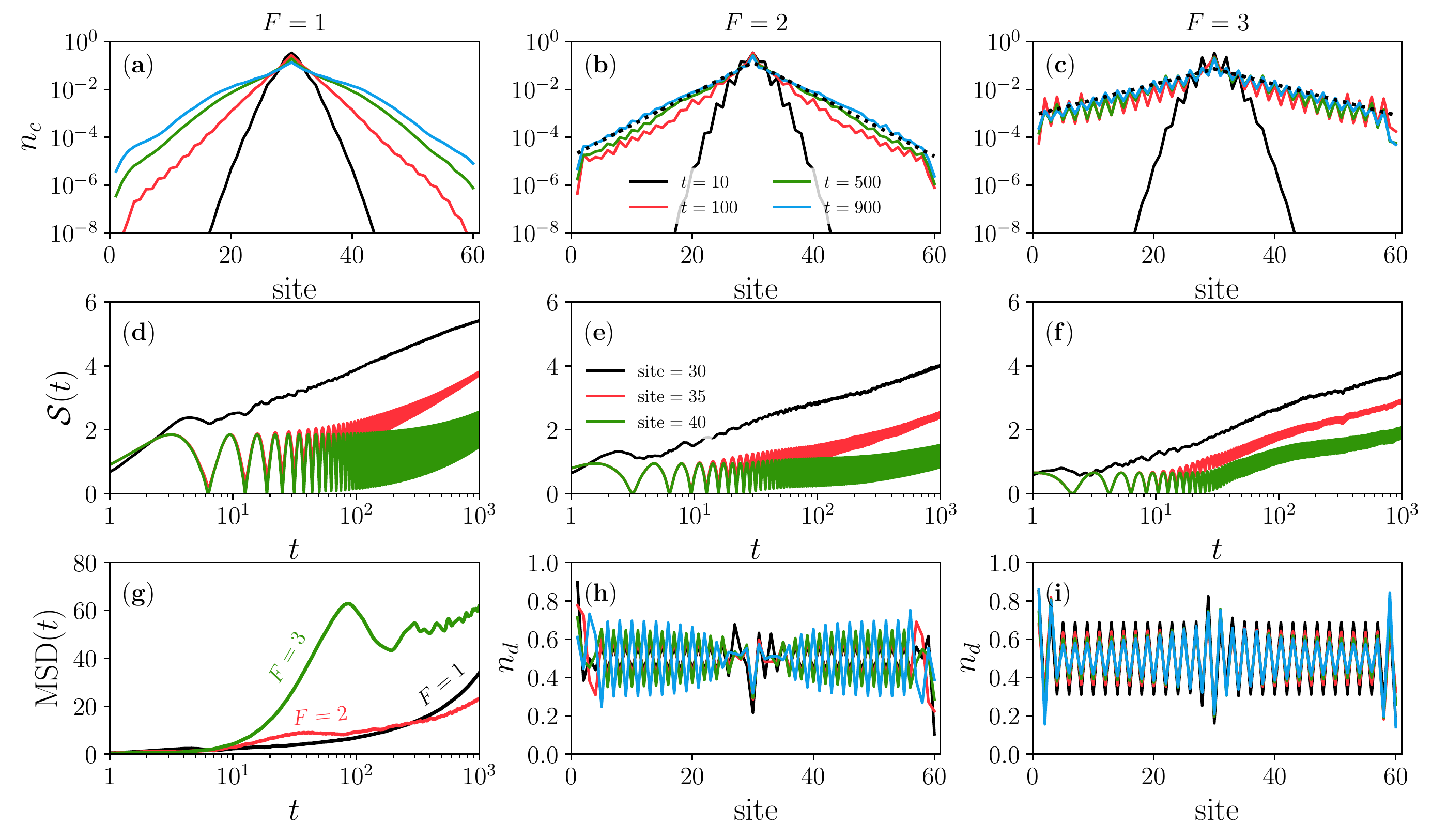}
  \caption{Time dynamics for half filling by $d$-particles and $U=12$ case for $F=1,2,3$ as indicated in the panels. Top row (a-c) shows the density of $c$ particles while medium row (d-f) the entanglement entropy at different cuts: in the middle (black) 5 sites to the right (red) and at site 40 (10 sites from the center, green line). Panel g) shows MSD for F=1 (black) F=2 (red) and F=3(green) while panels h) and i) - the distribution of d-particles for F=2 and F=3 for the same times as in (b) and (c). \new{Additionally, panels (b) and (c) show exponential fits to the density $n_c=A\exp(-B|i-i_0|)$ with $B=0.30$ for $F=2$ and $B=0.15$ for $F=3$.}
  }\label{fig:F123}
\end{figure*} 

The motion of $d$ particles feeling the tilted lattice is also interesting.
The initial density wave distribution of tilted particles significantly evolves in time. One can observe two regions. In the central zone, around the original position of the $c$-{boson},  the density of $d$-bosons seems to reach the mean 1/3 value apparently loosing any memory of initial occupations. 
However, moving outside, closer to the edges,  the partial memory of the initial distribution persists (except very close to the edges). Comparing data at different times one may observe that the density around the center is almost stationary while in the outside region, Bloch-like oscillations are still visible as could be inferred from panel Fig.~\ref{fig:L60dens}b). This is better visualized by the time dependent correlation function, Eq.~\eqref{corr}. If only \new{the} edges are removed by taking $L_b=3$, Bloch like oscillations of $C_d(t,3)$
are quite pronounced. On the other hand for $L_b=25$, when we measure the correlation function on ten central sites only, the oscillations are to a large extent suppressed as shown in panel Fig.~\ref{fig:L60dens}d).
Let us 
observe also that for the relatively large system, $L=60$,   no apparent asymmetry and accumulation of particles at one edge occurs.

Figure~\ref{fig:L60ent} presents the time dynamics of the entanglement entropies at different cuts of the system. Black lines correspond to the central cut at the initial position of the impurity, and red (green) lines correspond to cuts 5 (10) sites to the right, respectively. One clearly observes the delay of the entropy growth, the perturbation induced by the impurity spreads with a finite velocity across the system. Observe that while \new{the} configurational part of the entropy grows smoothly the number entropy reveals strong oscillations corresponding to fast Bloch density oscillations. The growth of the entropy is quite significant  pointing out towards \new{the} spreading of information across the whole system.

The corresponding scenario at $F=3$ for $L=60$ is quite similar to the story for small sizes as shown in Fig.~\ref{fig:L60F3}. The $c$ particle spreads over the whole system revealing, however, traces of localization at the center. Contrary to $F=2$ case, the density shows a ``teeth'' structure as the $c$-particle \new{avoids} sites where $d$-particles are present. \new{Still one may try to fit the exponentially decaying function, $n(i)=A\exp(-B|i-i_0|)$, centered at the initial $c$-particle position to the long time, $t=900$, density.
The result, on one hand, seems to reproduce the envelope quite well, on the other the numerical density shows small upward trend close to the edges of the chain. Also the resulting ``localization length'', $1/B\approx 8$ seems quite large showing again that the finite size of the system may affect the distribution. The $d$-particles, on the other hand,}   are almost \new{frozen} in their original positions thus intuition coming from the KP potential seems again applicable. The MSD (panel c) reveals a characteristic small maximum in its initial spread. The later growth as well as the growth of the entanglement entropy (not shown) \new{appears} to be logarithmic.
\new{Putting all these observations together, this case  seems to be very close to being localized. We have no evidence, up to the longest time studied by TDVP, of delocalization of $d$-particles, also the practical freezing of $c$-particle distribution and slow growth of MSD and $S$
points towards this interpretation which cannot be made definitive due to the finite times studied.}

Let us now move to \new{the} arguably more interesting $d$-particles half-filling case for which more clear signatures of MBL have been observed for small systems. It is particularly interesting in view of recent study of size dependence of MBL and of its thermodynamic limit. Several works observed a clear shift of MBL threshold with the system size, see e.g. \cite{Suntajs20,Suntajs20e,Sierant20b,Sierant20p}. It is interesting to see whether the same behavior occurs for our impurity.

\begin{figure}
    \includegraphics[width = 0.49\textwidth]{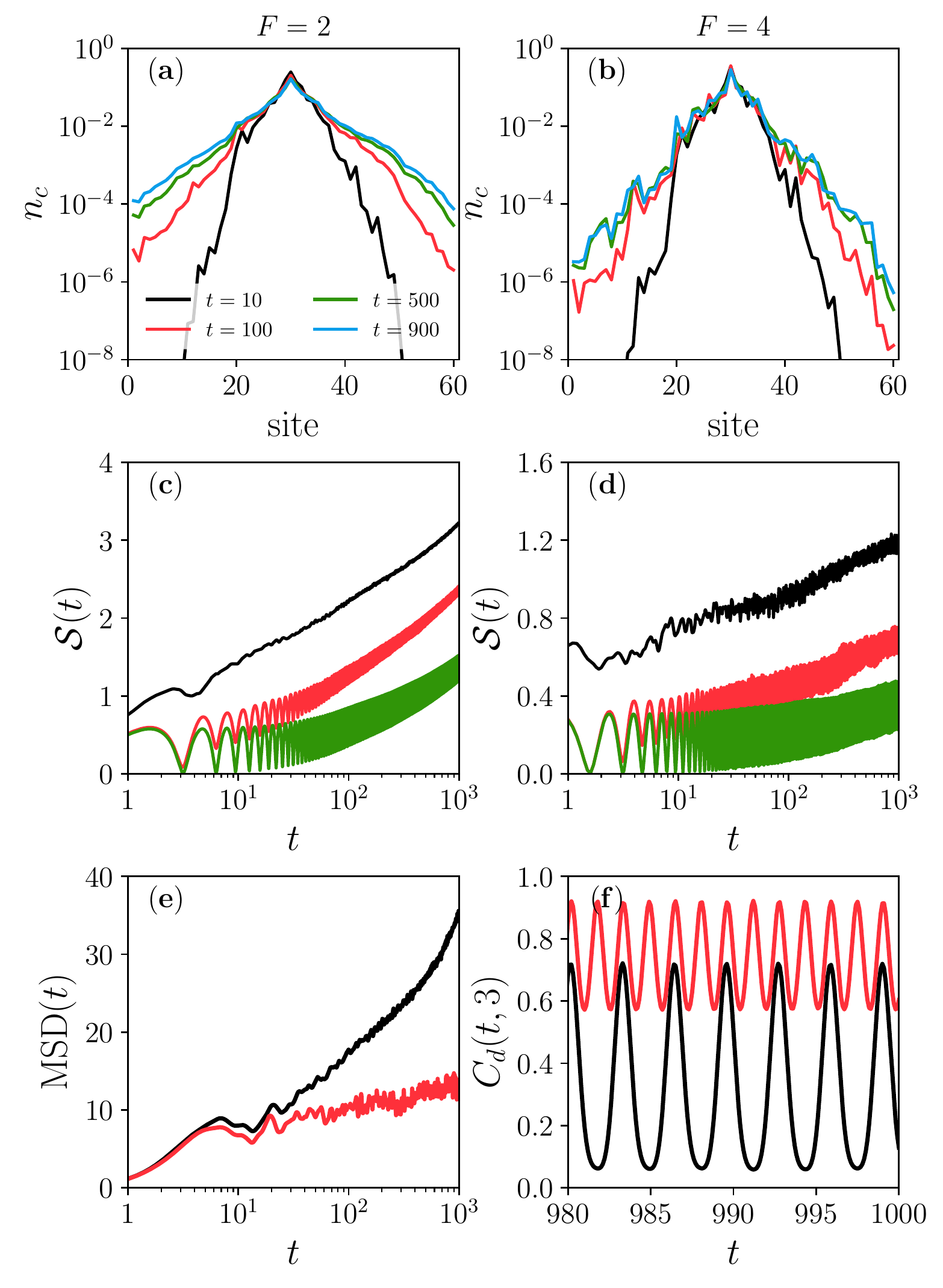}
  \caption{Time dynamics for 1/3 filling (randomly distributed $d$-bosons), $U=12$ case for $F=2$ (left column) and $F=4$ (right column). Strong exponential-like localization of $c$-boson (top row) is observed. It is accompanied by a log growth of the entanglement entropy (d) and a similar growth of MSD [panel (e) red line] for $F=4$. For $F=2$ entanglement entropy grows slightly faster than logarithmically [panel (c)], also MSD grows faster indicating lack of localization in that case.
  As in the previous figure we show the entropies at three different bonds: in the middle position $i_0$ and at $i_0+5$. $i_0+10$ to visualize spreading of entanglement in the system.
  Panel (f) shows correlations of d-bosons which remain very large and reveal fast Bloch oscillations. 
  }\label{fig:F24}
\end{figure} 

In the top row of Fig.~\ref{fig:F123} the gradual spreading of the $c$ particle may be seen. \new{The fast initial expansion slows down considerably for $t>100$, practically freezing for $F=2$ or $F=3$. The $F=2$ case is particularly interesting with $n_c$ revealing an} exponential-like envelope that is as low as $10^{-5}$ at the edges. For smaller $F=1$ the spreading is still significant even at late times while for $F=3$, localization is clearly weaker reaching around $10^{-3}$ at the edges. This is reflected in  
Fig.~\ref{fig:F123}g) that shows the time dynamics of MSD. Its growth is slowest for $F=2$ notice also the small (keeping in mind that we consider $L=60$) value of MSD even at $t=1000$. For $F=1$ MSD grows faster while for $F=3$ it is much larger. The entropy of entanglement seems to grow logarithmically for all $F$ values considered. Still neither $F=1$ nor $F=3$ case seems to show features typical for MBL.
Thus the slow, logarithmic-like entanglement entropy growth should not be considered as a proof of MBL.

\section{The role of positional disorder}
\label{RA}

Up till now we considered the situation when the $d$-particles, feeling the tilted lattice potential
were distributed regularly in the density wave pattern, either $|1,0,1,0..\rangle$ for half-filling or,
at the beginning for 1/3-filling  $|0,1,0,0,1,0..\rangle$. In both cases for a strong tilt this structure of the $d$-particles distribution imitated, to some extend, the KP potential for $c$ particle
due to interactions. Instead of the regular density wave pattern, one may consider a random distribution of $d$ particles over the sites. In this way, making $d$-particles for a moment immobile, 
we create the position-disordered potential for $c$ particle which should then become Anderson-localized. Of course the tunneling of $d$-particles destroys this picture but if the tilt is sufficiently strong, the Stark localization of $d$-particles prevents their tunneling so \new{the} $c$ particle sees \new{something approximating} a disordered potential. Do we observe MBL in this case?

Indeed this seems to be  the case for \new{a} strong tilt, $F=4$, as shown in Fig.~\ref{fig:F24}. We present here results obtained after averaging over 24 realizations of distribution of $d$ particles for 1/3-filling, i.e., for random distribution of 20 $d$ - particles on 60 sites. \new{For all the realizations}, the $c$ particle was initially \new{located} at the center at site number 30. We have taken only those random realizations in which this middle site was \new{not occupied by a $d$ particle.}
\new{This} enables comparison with density wave scenario where also the central site was \new{not occupied by a $d$ particle}.

For $F=4$ the situation seems very clear.  The entanglement entropy reaches a little above unity at final times $t=1000$ revealing a clear logarithmic growth. 
Its
oscillatory character indicates also that the entanglement entropy is dominated by the number entropy.
Similar is the fate of MSD which grows logarithmically. For $F=2$ it seems, at the first glance, that entropy also grows logarithmically, a careful inspection at later times reveals a slightly faster growth. This signature of a long time delocalization is magnified by the behaviour of MSD which seems to grow much faster at late times.
The correlation of $d$ particles reveals very smooth Bloch-like oscillations, regardless of the random particle distribution which apparently does not affect the regularity of these oscillations.

\section{Conclusions and perspectives.} 
\label{CO}

We have analyzed in detail the \new{behavior} of an impurity interacting with other particles that alone would be Wannier-Stark localized due to the lattice tilt.
{On purpose, we considered a very similar situation to that studied in recent works \cite{Brighi21a,Brighi21b,Brighi23,Sierant23i} where the background particles were Anderson localized.} 

Analysis of small system sizes\new{, performed by Hamiltonian exponentiation technique \cite{Weinberg17,Weinberg19},} allowed us to study long-time dynamics. We have shown that even for strong interactions, comparable to that studied in earlier works on random systems  \cite{Brighi21a,Brighi21b,Brighi23,Sierant23i}, for low density of $d$-particles ({one}-third filling) the resulting dynamics while clearly non-ergodic, could not be considered as
truly many-body localized. On the other hand, for the optimal one-half density, when the interactions are effectively maximized, clear manifestations of MBL were observed e.g. via the slow logarithmic growth of the entanglement entropy or of the mean squared deviation of $c$-particle distribution.

We have observed an interesting effect absent in random-localized case and present here for  significant inter-species interaction strength, $U$, and  a sufficiently large tilt, $F$. Such a tilt makes the Wannier-Stark localization length (and the amplitude of Bloch oscillations) very small, much smaller than the lattice spacings. This pins down $d$-particles
to their original positions and they generate an effective potential for $c$ particles.
For regular distribution of $d$-particles the resulting  potential is periodic resembling Kronig-Penney potential
and $c$ particles spreads in it. Interestingly, at later times the distribution reveals some signature of stabilization (with large fluctuations of MSD) accompanied by a logarithmic in time entropy growth. This suggests localization.

The results for small systems were fully confirmed with tensor network TDVP numerical calculations that could be carried out up to 1000 tunneling times (i.e. well below current experimental capabilities \cite{Scherg21}. The fact that the calculations could reach such long times indicates limited entanglement build-up and localized character of observed distributions. 

The situation dramatically changes for \new{a} random distribution of $d$-particles in the tilted potential.
Then for large tilt, they form a random potential due to their random positions. The coupled evolution of the full system shows then much stronger signatures of MBL with much slower entanglement entropy growth as well as an extremely slow spread of the $c$ particle density. Clearly, such a combination of a strong tilt and random particle distribution leads to robust many-body localization.

This suggests that even in the absence of a tilt one could observe MBL more clearly in the random case 
as the one studied in \cite{Brighi21a,Brighi21b,Brighi23,Sierant23i}. One need then sufficiently strong disorder to have the Anderson localization length of the order of the lattice site spacing
for $d$ particles but, importantly, they should be randomly and not regularly (as in \cite{Brighi21a,Brighi21b,Brighi23,Sierant23i}) distributed. A study in this direction is in progress.

\begin{acknowledgments}
 We are grateful to Krzysztof Sacha and Piotr Sierant for enlightening discussions, \new{to Oliver Readon-Smith for critical remarks} as well as to Adith Sai Aramthottil for useful hints on numerics. JZ is supported by  the National Science Centre (Poland) under grant OPUS18 2019/35/B/ST2/00034. The work of PRNF was funded by the National Science Centre, Poland, project 2021/03/Y/ST2/00186 within the QuantERA II Programme that has received funding from the European Union Horizon 2020 research and innovation programme under Grant agreement No 101017733.  We gratefully acknowledge Polish high-performance computing infrastructure PLGrid (HPC Centers: ACK Cyfronet AGH) for providing computer facilities and support within computational grant no. PLG/2022/015613. Views and opinions expressed in this work are, however, those of the author(s) only and do not necessarily reflect those of the European Union, European Climate, Infrastructure and Environment Executive Agency (CINEA), nor any other granting authority. Neither the European Union nor any granting authority can be held responsible for them. Authors declare that they did not involve any artificial intelligence tool in writing this manuscript. For the purpose of
Open Access, the authors applied a CC-BY public copyright
licence to any Author Accepted Manuscript (AAM) version arising from this submission.

\end{acknowledgments}
\section{Appendix}
\label{APP}
We provide here more details on the TDVP algorithm and its application to the problem studied. The foundation of the method as well as details of its implementation may be found, e.g., in a recent review \cite{Paeckel19}. We mention that our implementation has been tested already in several numerical studies of dynamics in many-body systems, in particular those related to MBL \cite{Chanda20,Chanda20c,Chanda20m,Chanda20t,Yao20b,Yao21,Yao21a,Sierant23i}. 

\begin{figure}
    \centering
    \includegraphics[width = 0.49 \textwidth]{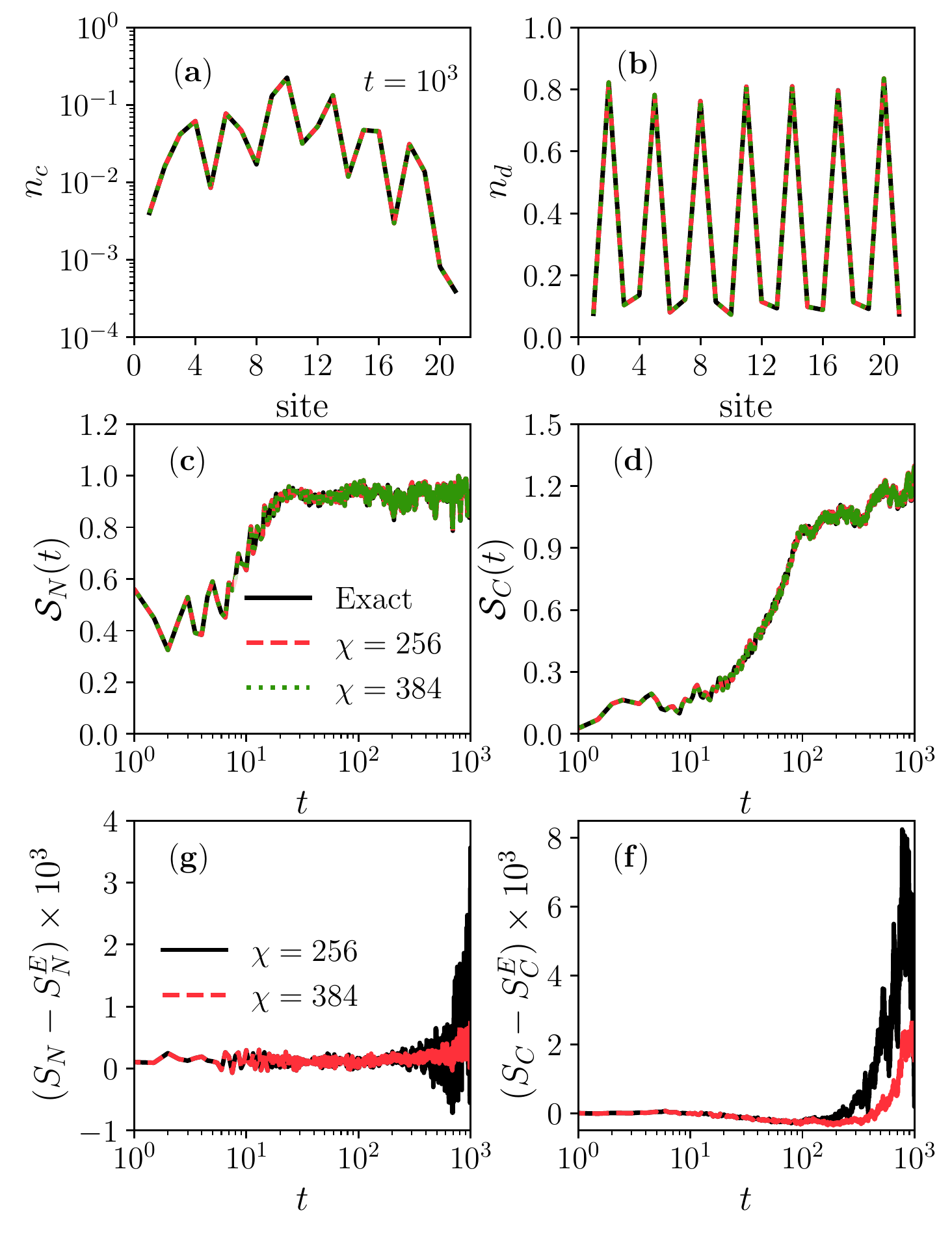}
    \caption{Comparison of  time dynamics obtained with Hamiltonian exponentiation with results of TDVP algorithm for $L=21$, $F=4$, $U=12$. Top row shows the occupations of sites for the impurity (left) and $d$-particles. The agreement is spectacular. \new{The} middle row shows the number and confugurational parts of the entanglement entropy. \new{Their errors (note the change of the vertical scale) are shown in the bottom row.}
    }
    \label{fig:app}
\end{figure}

In numerical studies performed in this study we use a hybryd version of TDVP \cite{Paeckel19,Goto19}. At the initial stage it uses a two-site version of TDVP. The initial state is represented as a matrix product state (since we start the time evolution from \new{product} states this is straightforward).
During the time evolution the auxiliary bond dimension grows reflecting the build-up of the entanglement in the system.
We assume some maximal allowed bond dimension $\chi$ and the evolution is followed with the two-site version until this value of $\chi$ is reached. Then we switch to the single-site version. That allows us to partially avoid errors related to the truncation of the singular values in the two-site version
\cite{Paeckel19,Goto19}. Controlling the errors related to a necessary restriction of the Hilbert space dimension as well as comparing the results for different $\chi$ allows us to assess the convergence of the results.

A comparison of the TDVP performance with the ``exact'' results may be carried out for small system sizes only. Fig.~\ref{fig:app} shows such a comparison for the $F=4$, $U=12$ system for $L=21$. In  the whole time interval studied the agreeement is highly satisfactory, the curves resulting from the exact and TDVP evolution practically coincide. In fact, the occupation of sites is not a very sensitive measure
of this agreement, more details are provided by comparison of time dynamics of entropies. As shown in the bottom row of Fig.~\ref{fig:app} some differences between exact entropies and those obtained using TDVP are visible for late times. 
Still even for the largest time considered, $t=1000$ the errors
for $\chi=256$ are below 1\%, being about 6 times smaller for $\chi=384$. For larger system sizes no comparison with ``quasi-exact'' time dynamics is possible so we could compare the results of simulations for different $\chi$ only. A comparison with smaller $\chi$ results showed small differences (for entropies) at large times, differences that did not affect the conclusions presented.

We must note, at the same time, that large values of $U$ and $F$ made the TDVP algorithm to be very costly in the execution time. Some of the data presented required about two weeks of single processor time. Our version of TDVP uses \new{a} typical for standard density matrix renormalization group (DMRG) technique of sweeping along the chain that prevented us from an otherwise possible parallelization of the algorithm.

\normalem
%


\end{document}